\documentclass[linenumbers]{aastex631}

\newcommand\xmmsrc{2XMM J104608.7$-$594306}

\begin{document}

\title{A coherent radio burst from an X-ray neutron star in the Carina Nebula}

\author[0000-0002-8043-6909]{K. M. Rajwade}
\affiliation{Astrophysics, University of Oxford \\ 
Denys Wilkinson Building, Keble Road \\
Oxford OX1 3RH, UK}

\author[0000-0002-4842-7076]{J. Tian}
\affiliation{Jodrell Bank Centre for Astrophysics,\\
University of Manchester, Oxford road, Manchester, \\
M13 9PL, UK}

\author[0000-0002-7991-028X]{G. Younes}
\affiliation{Astrophysics Science Division, \\ 
NASA Goddard Space Flight Center, Garden street 8800 \\
Greenbelt Road, Greenbelt, MD 20771, USA}

\author[0000-0003-2317-9747]{B. Posselt}
\affiliation{Astrophysics, University of Oxford \\ 
Denys Wilkinson Building, Keble Road \\
Oxford OX1 3RH, UK}

\author[0000-0001-9242-7041]{B. Stappers}
\affiliation{Jodrell Bank Centre for Astrophysics,\\
University of Manchester, Oxford road, Manchester, \\
M13 9PL, UK}

\author[0000-0002-9249-0515]{Z. Wadiasingh}
\affiliation{Astrophysics Science Division, \\ 
NASA Goddard Space Flight Center, Garden street 8800 \\
Greenbelt Road, Greenbelt, MD 20771, USA}

\author{E. D. Barr}
\affiliation{Max-Planck-Institut f{\"u}r Radioastronomie \\
Auf dem H{\"u}gel, D-53121 Bonn, Germany}

\author{M. C. Bezuidenhout}
\affiliation{Department of Mathematical Sciences, University of South Africa \\ 
Cnr Christiaan de Wet Rd and Pioneer Avenue, Florida Park, 1709 \\
Roodepoort, South Africa}

\author[0000-0002-4079-4648]{M. Caleb}
\affiliation{Sydney Institute for Astronomy, School of Physics, The University of Sydney \\
 Sydney 2006, NSW, Australia}
 \affiliation{ARC Centre of Excellence for Gravitational Wave Discovery (OzGrav) \\
 Hawthorn 3122, Victoria, Australia}

\author[0000-0002-6658-2811]{F. Jankowski}
\affiliation{LPC2E, OSUC, Univ Orleans \\
 CNRS, CNES, Observatoire de Paris, F-45071 Orleans, France}

\author[0000-0002-4175-2271]{M. Kramer}
\affiliation{Max-Planck-Institut f{\"u}r Radioastronomie \\
Auf dem H{\"u}gel, D-53121 Bonn, Germany}

\author[0000-0002-4357-8027]{I. Pastor-Marazuela}
\affiliation{Jodrell Bank Centre for Astrophysics,\\
University of Manchester, Oxford road, Manchester, \\
M13 9PL, UK}

\author{M. Surnis}
\affiliation{Department of Physics, Indian Institute of Science Education and Research Bhopal \\ Bhopal Bypass Road, Bhauri Bhopal 462 066, Madhya Pradesh India}



\begin{abstract}

The neutron star zoo comprises several sub-populations that range from energetic magnetars and thermally emitting X-ray neutron stars to radio-emitting pulsars. Despite studies over the last five decades, it has been challenging to obtain a clear physical link between the various populations of neutron stars, vital to constrain their formation and evolutionary pathways. Here we report the detection of a burst of coherent radio emission from a known radio-quiet, thermally emitting neutron star~\xmmsrc~in the Carina Nebula. The burst has a distinctive  sharp rise followed by a decay made up of multiple components which is unlike anything seen from other radio emitting neutron stars. It suggests an episodic event from the neutron star surface akin to transient radio emission seen from magnetars. The radio burst confirms that the X-ray source is a neutron star and suggests a new link between these apparently radio-quiet X-ray emitting sources and other transient or persistent radio emitting neutron stars. It also suggests that a common physical mechanism for emission might operate over a range of magnetic field strengths and neutron star ages. We propose that \xmmsrc\, straddles the boundary between young, energetic neutron stars and their evolved radio-emitting cousins and may bridge these two populations. The detection of such a radio burst also shows that other radio-quiet neutron stars may also emit such sporadic radio emission that has been missed by previous radio surveys and highlights the need for regular monitoring of this unique sub-population of neutron stars.

\end{abstract}



\section{Introduction} \label{sec:intro}

Many questions still remain unanswered about the evolution of neutron stars in our Galaxy. The problem is exacerbated by the large diversity observed in the electromagnetic radiation from these compact remnants. Among the neutron star zoo are highly magnetized neutron stars (magnetars), X-ray Dim Isolated Neutron Stars (XDINSs) and Compact Central Objects (CCOs) whose salient properties differ from the majority of the neutron star population typically seen as rotationally powered radio neutron stars (pulsars). Magnetars are characterised by episodic outbursts visible in X-rays triggering radio emission in some cases that lasts only for a few years~\citep{kaspi2017,camilo2006}. The emission seen from magnetars is believed to be powered by the large magnetic field energy possessed by these objects.

XDINSs on the other hand are radio-quiet neutron stars and are characterized by their extremely low X-ray luminosity, and X-ray pulsed fraction when compared to the typical thermal luminosity of neutron stars in the Galaxy. This extreme dimness in X-rays challenges our understanding of neutron star cooling mechanisms and emission processes, raising questions about the physical parameters that influence the luminosity and thermal evolution~\citep{vigano2013}. Unraveling the mysteries of XDINSs is crucial not only for the understanding of the diverse outcomes of stellar evolution but also for probing the physical conditions under which neutron stars transition from active X-ray pulsars to low-luminosity objects~\citep{beniamini2023}.

CCOs are compact objects in the central regions of supernova remnants. These neutron stars are young and hotter than XDINSs. Intriguingly, none have been detected at radio wavelengths~\citep{deluca2017}. Most CCOs show X-ray pulsations with a significant pulsed X-ray flux and fast spins that one would expect from a newly born neutron star but unlike a nascent neutron star, the estimated magnetic fields of CCOs are 10$^{11}$~G. This is lower by at least an order of magnitude with respect to the bulk of the young neutron star population, including neutron stars with an age of the order of 10$^{5}$~years. All these attributes makes CCOs stand out from the neutron star zoo and highlights the importance of identifying their connection with the other classes of neutron stars. The connection between CCOs and magnetars comes from a unique neutron star called 1E~161348-5055~\citep{rea2016} that has a claimed period of 6.7 hours and shows magnetar like outbursts. The lack of radio pulsations from CCOs can be attributed to unfavourable beaming or the lack of radio emission due to their lower magnetic fields~\citep{Luo2015}.

The connection between XDINSs, CCOs, magnetars and pulsars have been investigated in the past. Two key questions that arise are: 1) Are magnetars/CCOs, XDINSs and pulsars part of the same evolutionary phase of isolated neutron stars and 2) What is the origin of the lack of radio-emission from these source? Of the 7 confirmed XDINSs~\citep{haberl2007} and 10 confirmed CCOs~\citep{deluca2017} known so far, apart from a few instances of claimed radio detections~\citep{malofeev2006}, none have shown confirmed and persistent coherent radio emission in spite of several searches for the same~\citep{kondratiev2009, ipm2023b,lu2024}. The remarkable similarity between the X-ray properties of XDINSs and a persistent radio pulsar PSR~J0726$-$2612~\citep{rigoselli2019} has led astronomers to suggest that the radio-quietness of XDINSs and CCOs is mostly attributed to beaming and the combined evolution of the inclination angle of the magnetic axis and the age of the XDINSs could explain this dearth. 

Another aspect that one needs to consider for the radio quietness of XDINSs/ CCOs is intermittency and variability.  A number of radio-loud neutron stars and rotating radio transients (RRATs)~\citep{mclaughlin2006} show extreme intermittency and variability in their radio emission. One extreme example of this is the population of intermittent pulsars where the radio emission mechanism switches off for years~\citep{kramer2006}. The detection of bright, sporadic radio bursts from a Galactic Magentar SGR~J1935+2154 clearly showed astronomers that radio emission in neutron stars can turn on only for a short time (~single rotation)~\citep{SGR1935chime, bochenek2020}. This highlights the fact that finding neutron stars where the radio emission mechanism is inactive and can turn on briefly could provide the missing piece of the puzzle that signals the transitionary phase in the evolution of a neutron star between the various sub-populations. In this work, we report the discovery of a coherent radio burst associated with a radio-quiet X-ray bright neutron star~\xmmsrc~that provides evidence for such a link. The paper is organized as follows: in section 2, we describe the discovery. In section 3, we discuss the radio properties of the burst. In section 4, we present the results from our analysis of archival and follow-up X-ray observations. We discuss the implication of this discovery in section 5 and present our summary in section 6.

\section{Discovery} \label{sec:obs}

\begin{figure*}
    \centering
    \includegraphics[width=0.9\linewidth]{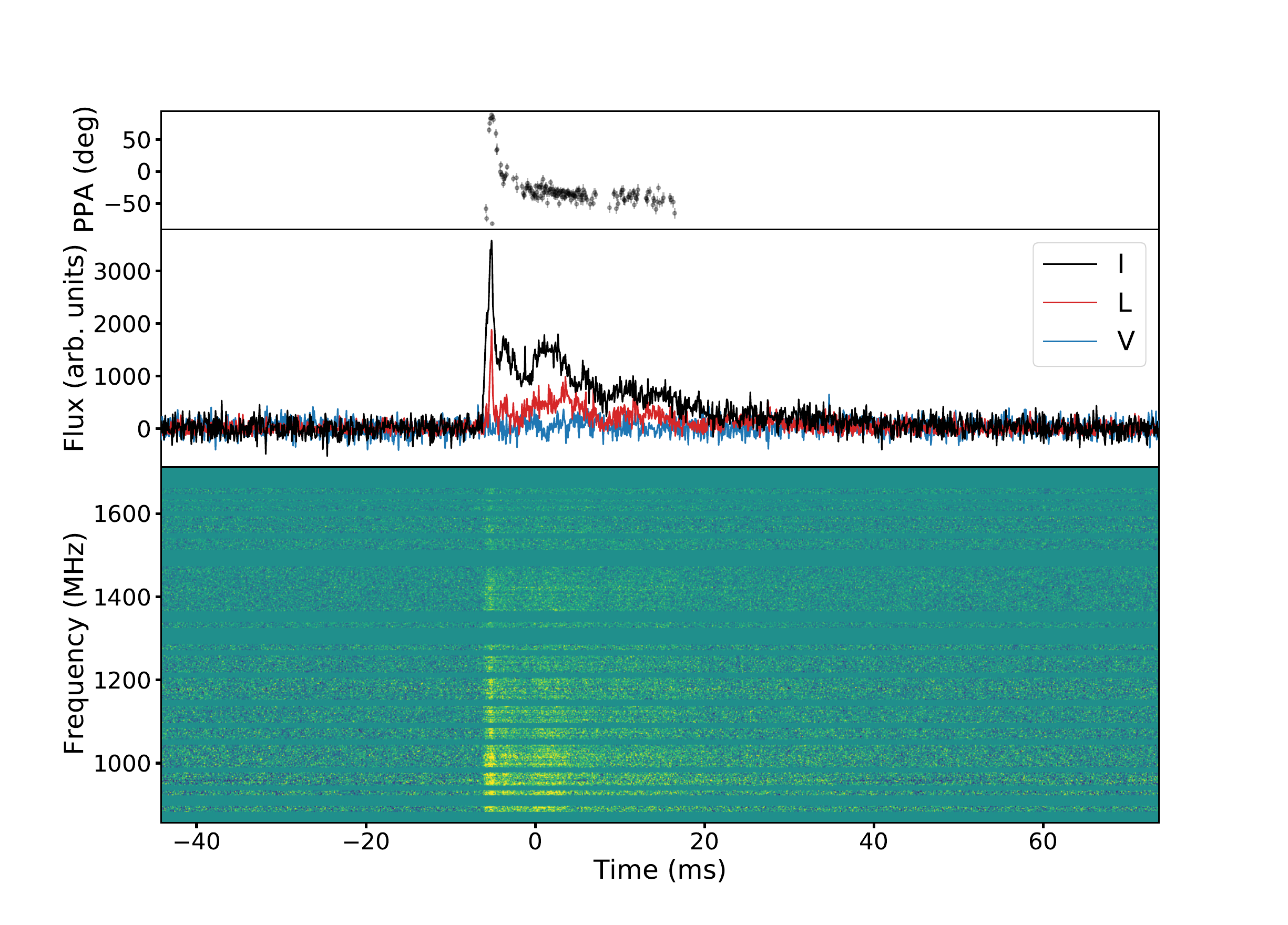}
    \caption{Radio burst from~\xmmsrc~showing linear polarization (red), circular polarization (blue) and the polarization position angle (black points). The pulse has been coherently dedispersed at the DM that maximises the S/N. The bottom panel shows a dynamic spectrum of the total intensity as a function of time and observing frequency, with horizontal excisions for RFI.}
    \label{fig:pulseprofile}
\end{figure*}

We discovered a single radio burst with the MeerKAT radio
telescope during an open-time science observation (Proposal
ID: SCI-20230907-MB-01) by the MeerTRAP instrument, a commensal fast radio transient detection system~\citep{rajwade2022} on UTC 2024-02-03 06:55:16.738. Notably, the burst 
exhibited peculiar temporal morphology that resembles the
fast rise, exponential decay (see 
Figure~\ref{fig:pulseprofile}), typically only seen in 
extragalactic gamma-ray bursts, X-ray flares from 
thermonuclear bursts in accreting neutron stars, and a few 
X-ray bursts from magnetars~\citep{russell2024, collazzi2011}. The burst was polarized with a linear polarization fraction of $\sim30\%$ and no evidence of circular polarization.  A search in archival MeerKAT observations at the same observing frequency (1.284~GHz) of the same field revealed no other bursts. The real-time detection triggered the storage of complex channelized voltages from all available MeerKAT antennas at the time of the burst~\citep{rajwade2024}. These data were used to create images at a 153~$\mu$s time resolution to localise the burst to RA (J2000) = 
$10^{h}46^{m}08.6^{s}$(1.8) and DEC (J2000) = $-59^{d}43^{m}06.8^{s}$(1.4) (see Appendix~\ref{sec:loc} and Figure~\ref{fig:localisation}). This position is consistent 
to within $1\arcsec$ of the position of the known 
thermally X-ray emitting neutron star~\xmmsrc~in the Carina Nebula region that is considered to share characteristics with XDINSs and CCOs~\citep{pires2015}.
At this offset, the chance coincidence of spatial association can be ruled out at a 99.9$\%$ confidence level (see Appendix~\ref{sec:loc}), confirming that the radio burst originated from~\xmmsrc. Thus, the radio burst corroborates the neutron star nature of~\xmmsrc. The dispersion measure (DM) of the burst (98.6$\pm$3~pc~cm$^{-3}$) translates to a distance of 1.5--2.3~kpc based on the YMW16~\citep{ymw16} and NE2001~\citep{ne2001} Galactic electron density models respectively. This distance range is consistent with the reported distance for~\xmmsrc~\citep{pires2012}. The radio luminosity of the burst (see Appendix~\ref{sec:rl}) is $\sim$4.6$\times$10$^{29}$~ergs~s$^{-1}$ assuming isotropic emission and a distance of 1.5~kpc. All the properties of the burst are shown in Table~\ref{tab:mtp9g_prop}.

\begin{figure}
    \centering
        \includegraphics[width=0.9\textwidth]{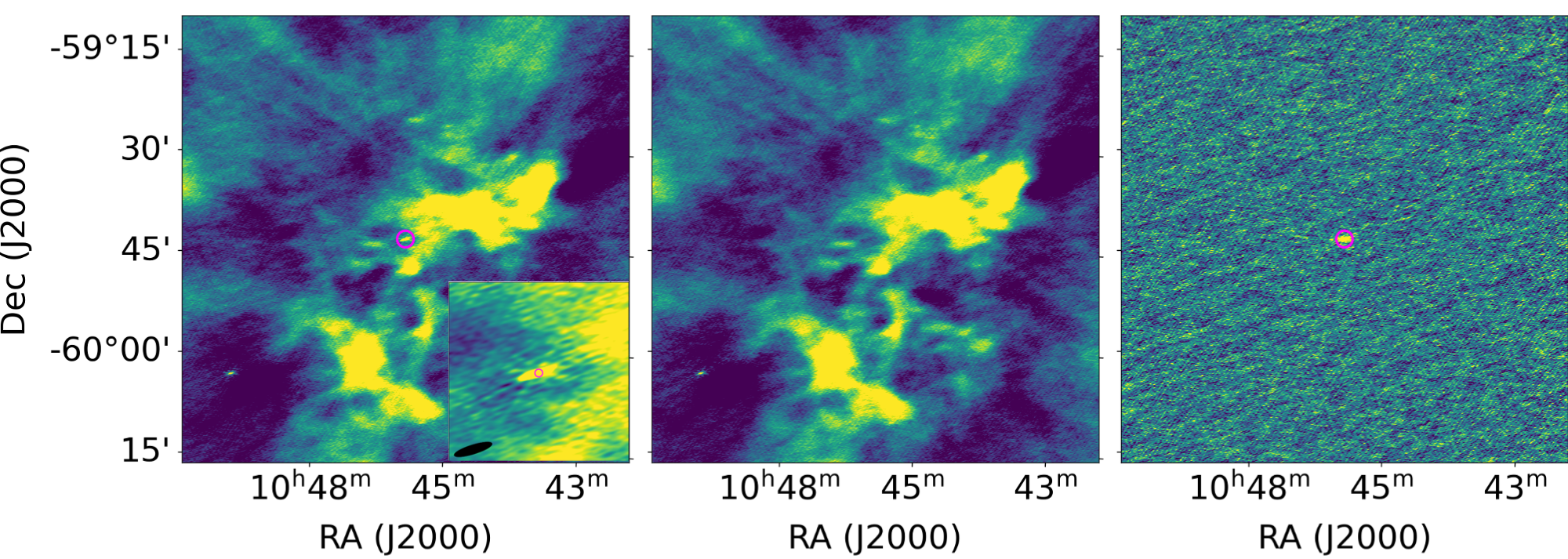}
    \caption{Images of the field of \xmmsrc\, integrated over the duration of the radio burst (left) and before the burst detection (centre) and the difference between the two images (right). The magenta circle marks the transient source identified at the time of the burst detection, and the inset at the bottom right corner shows a zoomed in view to display the position (in red) of the XDIN candidate \xmmsrc~on top of the transient source. The images have a synthesised beam size of $43"\times9"$. The black ellipse in the inset at the bottom right corner of the left panel shows the synthesized beam.
}
\label{fig:localisation}
\end{figure}

\section{Burst properties}

The radio burst from \xmmsrc~exhibits a drastic sharp rise and a slowly decaying profile, and at least 8 distinct emission components (see Appendix~\ref{sec:qpsearch}). The temporal structure in the burst resembles quasi-periodic micro-structure seen in single pulses from radio-emitting neutron stars~\citep{ipm2023, kramer2024}. Such structure may point to a disturbance in the magnetosphere precipitated by activity in the neutron star crust~\citep{wadiasingh2020b}. A search for Hz--kHz periods in burst morphology did not result in any significant detections above 2.1-$\sigma$ (see Appendix~\ref{sec:qpsearch}). To characterize the peculiar morphology, we compute the rise time of the leading edge. To do this, we assume the emission starts from the time sample where the intensity is greater than 3 times the standard deviation of the noise in the baseline.  The rise time is defined as the time between this time sample and the peak of the leading component. Following this method, we compute a rise time of 1.1~ms.This timescale is much shorter than the rise timescale seen in Type-I X-ray bursts and magnetar X-ray bursts~\citep{collazzi2011, russell2024} but comparable to narrow emission structures seen in radio pulses from neutron stars~\citep{kramer2024}. The temporal structure of the profile could also arise from interstellar scattering expected from plasma in the Galaxy~\citep{Oswald2021} but the brightest and narrowest components shows no evidence of scattering in the dynamic spectrum (Figure~\ref{fig:pulseprofile}). Hence, we do not consider scattering to be responsible for the observed burst shape and if there is scattering present in the radio emission, the timescale is less than the width of the narrowest component (1.7~ms).

The peculiar burst morphology warranted a further quantification of the observed microstructure. We first estimated the optimal number of emission components (see Appendix~\ref{sec:qpsearch}). In order to find the optimal number of components, we fit the burst with a combination of $N$ Gaussian functions. For every $N$, we computed the Akaike Information Criterion (AIC) and the Bayesian Information Criterion (BIC) from the best fit model. The optimal number of emission components were then chosen based on those that gave the lowest value of AIC and BIC. We found 8 to be the optimal number of components with the BIC value of 127.9 and the AIC value of 35.3. We do note that the number of components are also covariant with the time binning of the data as the noise in the dataset will affect the number of optimal components. The time binning was chosen as an optimal value that minimizes the trade-off between components merging into one another and the data becoming too noisy for a robust fit. Hence, we report the number of components as a lower limit.

\begin{table*}
    \centering
 \begin{tabular}{lcc}
    Parameter & Unit & Value \\
    \hline
    MJD &  & 60343.28838817  \\
    UTC &  & 2024-02-03 06:55:16.738  \\
    RA & (hms) & 10:46:08.6$\pm$1.8"  \\
    Dec & (dms) & $-$59:43:06.8$\pm$1.4"  \\
    $l$ & (deg) & 287.7337 \\
    $b$ & (deg) & $-$0.5962  \\
    Detection centre frequency & (MHz) & 1284  \\
    DM & (pc~cm$^{-3}$) & 98.41$\pm$3  \\
    DM distance (NE2001/YMW16) & (kpc) & 1.5/2.3 \\
    Detection S/N & & 21   \\
    Beamformed S/N & & 125  \\
    W$_{\text{50p}}^{\rm boxcar}$ & (ms) & $23.14\pm0.3$  \\
    W$_{\text{50p}}^{\rm fit}$ & (ms) & $51.39\pm5.44$  \\
    RM &(rad~m$^{-2}$) & $-$50.23$\pm$1.2 \\
    L/I & & 0.3$\pm$0.05\\
    \hline
    $S_{\text{peak}}$ & (Jy) & $0.30\pm0.03$  \\
    Brightness temperature & (K) & 10$^{16}$ -- 10$^{19}$ \\
    $F$ & (Jy ms) & $7.1\pm$0.9   \\
    L$_{\rm radio (1284 MHz)}$ & (ergs~s$^{-1}$) & 4.6$\times$10$^{29}$\\
    \hline
    \end{tabular}
    \caption{Various observed and measured properties for the radio pulse from \xmmsrc. We note that for the flux calculations, we use the optimally detected pulse width (W$_{\text{50p}}^{\rm boxcar}$) as it assumes that the area under the pulse is conserved. The burst was originally detected in the incoherent beam and the data were re-beamformed at the burst location, which yielded a much larger S/N. The brightness temperature range is based on the entire burst duration and the width of the shortest emission component.}
    \label{tab:mtp9g_prop}
\end{table*}

\section{Multi-wavelength follow-up}

\subsection{Murriyang observations}

We carried out a 1-hr follow-up observation of \xmmsrc\ using the ultra-wideband low (UWL) receiver~\citep{Hobbs2020} at the 64-m Murriyang (formerly known as Parkes) radio telescope on 2024 March 01 from 09:51 UTC to 10:51 UTC. The observation spans a frequency range of $0.7\text{--}4$\,GHz and has 1\,MHz wide channels, sampled at a time resolution of $256\,\mu$s. We performed a single pulse search on this dataset using the software {\sc TransientX}\footnote{\url{https://github.com/ypmen/TransientX}}~\citep{Men24} over a DM range of $80\text{--}120\,\text{pc}\,\text{cm}^{-3}$ and a maximum pulse width of 1\,s. This resulted in 785 candidates above the $7\sigma$ threshold. After visually inspecting their pulse profiles and dynamic spectra, we identified all the candidates as radio-frequency interference (RFI). We did not find any promising candidates. Given the UWL receiver on Parkes has a system equivalent flux density ranging from 33 to 72\,Jy depending on frequency \citep{Hobbs2020}, our non-detection corresponds to a fluence upper-limit of 0.28--0.62\,Jy\,ms assuming a burst width of 10\,ms and a spectral extent spanning the entire band.
We also performed a Fast Folding Algorithm (FFA; \citep{Morello20}) search over a period range between 10\,ms--100\,s and the nominal DM on the Parkes data. We found no periodic signals of astrophysical origin. In conclusion, we did not detect \xmmsrc\ in the Parkes data using both the single pulse search and the FFA search. We estimate the flux limit of our search to be $\sim10~\mu$Jy for a duty cycle of 1\%. 

\subsection{High-energy searches}
Single radio bursts have been observed from 
isolated highly magnetized neutron stars, with contemporaneous high-energy emission in some cases~\citep{SGR1935chime,israel2021}. Hence, we carried out a search for any associated X-ray and soft $\gamma$-ray emission consistent with the location of~\xmmsrc. We find no publicly-reported flare-like emission at the time of the \xmmsrc\ radio burst from any of the all-sky high-energy monitors such as {\it Swift}-BAT~\citep{krimm2013}, {\it Fermi}-GBM~\citep{meegan2009} and INTEGRAL~\citep{jensen2003}. \xmmsrc\ was visible to {\it Fermi}-GBM at the time of the burst, and we derive a $3\sigma$ upper-limit of $F_{\rm X} \leq5.0\times10^{-8}$~erg~s$^{-1}$~cm$^{-2}$ for a 0.2~s burst-like emission in the 8-100 keV energy range, and a thermal spectrum with a temperature of 10 keV, properties similar to magnetar-like emission. We also searched for sub-threshold magnetar-like bursts in {\it Fermi}-GBM in a 2-month interval around the time of the radio burst and we find no credible bursts consistent with the location of \xmmsrc. Additionally, no magnetar-like bursting activity has ever been reported from the direction of the source with {\it Swift}-BAT. Lastly, we searched for periodic pulsed emission in archival X-ray data taken by the XMM-Newton and NICER telescopes. A $3.9~\sigma$ (post-trial) candidate periodicity of 18.6\,ms had been reported previously for this source~\citep{pires2012}. A confirmation linking this candidate periodicity to neutron star's spin period has not yet been established, since subsequent XMM-Newton observations 
were not sensitive enough due to a shorter effective exposure time, higher background, and the very low reported pulsed fraction of $p_f =13.1 \pm 1.6$\% of the X-ray pulse at energies $0.36-2.25$\,keV \citep{pires2015}. We were also unable to establish a periodicity in these archival and current X-ray datasets and report an upper-limit on the pulsed fraction of $14$\% and $15$\% for the NICER and XMM-Newton datasets respectively in a similar energy range and a period range of 1~ms to 100~seconds. These limits are also not constraining to conclusively exclude the 18.6\,ms candidate. However, we do note that the total duration of the radio burst ($\sim$ 50~ms) is much larger than the putative periodicity making this period unlikely if the radio burst originates from the same region of the star.

\section{Discussion}
\subsection{The link between XDINSs, magnetars and FRBs}
Our observations suggest that \xmmsrc~is a notable evolutionary link between neutron star populations -- either between XDINSs, magnetars, and rotation-powered radio pulsars \citep[similar to PSR J0726--2612][]{rigoselli2019} or between the latter and CCOs. For instance, the inferred blackbody temperature of \xmmsrc~ from previous X-ray observations is similar to those of XDINSs. An age of $\sim 10^6$\,yr for a NS in the Carina Nebula \citep{Hamaguchi2009} is also broadly consistent with XDINSs. XDINSs have always stood out as a distinct class of middle-aged (age of a few hundred\,kyr) neutron stars because of their unusually high X-ray luminosities \citep[e.g.,][]{Kaspi2010,yoneyama2019} that could be associated with magnetars and other sub-populations. Despite striking differences in their absolute X-ray luminosity ranges, magnetars and XDINS have similar periods suggesting period-freezing from field decay \citep{2000ApJ...529L..29C}.  These diverse evolutionary paths likely arise from variations in the sources' initial dipole and crustal toroidal fields~\citep[][]{2012MNRAS.422.2878D,2019MNRAS.487.1426B,igoshev2025protoneutron}. 
If \xmmsrc~ is related to XDINSs, then the detection of its radio burst has established for the first time that these objects are capable of producing sporadic coherent radio emission similar to radio bursts seen from magnetars~\citep{bochenek2020, SGR1935chime} and other radio-emitting neutron stars.

\begin{figure}
    \centering
    \includegraphics[width=0.7\linewidth]{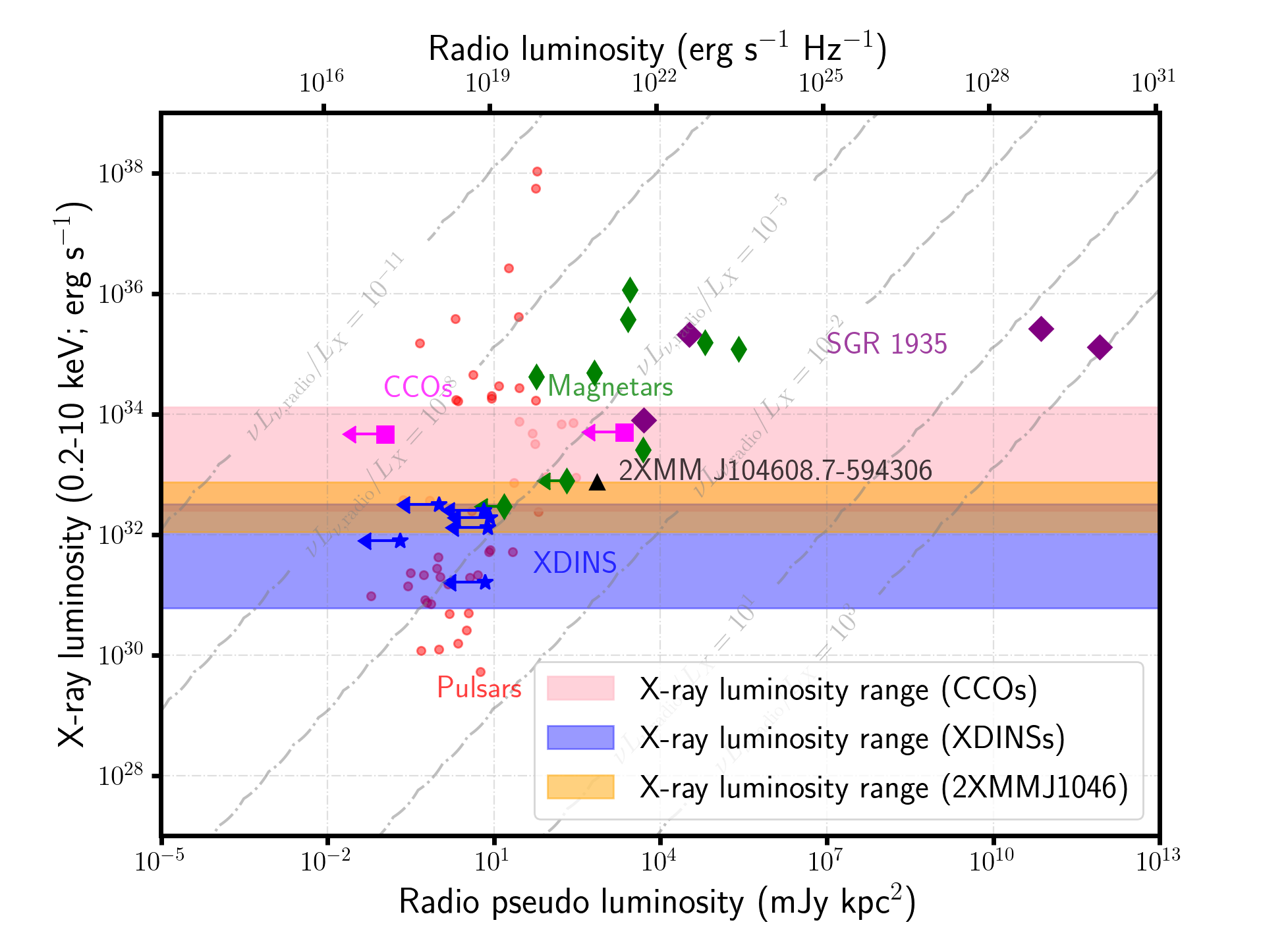}
 \caption{X-ray versus radio luminosity for various coherent radio-emitting sources.The shaded region shows the range of X-ray luminosities for XDINSs, CCOs and \xmmsrc. The dashed gray lines denote 
    constant ratios of the radio and X-ray luminosity (radio luminosity computed at 1.4~GHz). We note that the X-ray fluxes for pulsars, XDINSs and CCOs are the persistent X-ray fluxes reported in~\citet{Kaplan2009},~\cite{malov2018},~\cite{prinz2015},~\cite{haberl2007} and~\cite{deluca2017} for CCOs, while for magnetars, we use values from~\cite{israel2021},~\cite{esposito2020},~\cite{camilo2018} and~\cite{bochenek2020} and the references therein.
 For pulsar, we use fluxes based on the power-law fit in~\cite{prinz2015} in the 0.1--2~keV band and extrapolate them to 0.2-10~keV.
 The radio luminosities for pulsars and magnetars are the period-averaged radio luminosities~\citep{psrcat} i.e based on the continuum radio flux at 1.4~GHz. 
 Here, radio pseudo luminosity refers to the product of the average radio flux density and the square of the distance while the radio luminosity is calculated by multiplying the pseudo luminosity by 4$\pi$ (to denote isotropic radio luminosity). The values for SGR~J1935+2154 are taken from~\cite{bochenek2020} and~\cite{kirsten2021}.}
    \label{fig:phasespace}
\end{figure}

Figure~\ref{fig:phasespace} shows the location of~\xmmsrc~ in comparison to the different populations of neutron stars that emit at X-ray and radio wavelengths. The radio burst from the source has a similar luminosity to the average radio luminosity of radio-emitting magnetars and on average is brighter than the average luminosity of the normal radio pulsars. However, the burst is at least eight orders of magnitude fainter than the brightest radio burst from SGR~J1935+2154~\citep{2023SciA....9F6198Z}. We expect these bursts to be at a lower brightness due to a lower level of crust activity and magnetic free energy available to power bursts, or a lower pre-existing field twist and magnetospheric charge density  ~\citep{2019ApJ...879....4W,wadiasingh2020, cooper2024}. The X-ray luminosity of \xmmsrc~ when compared with other magnetars, XDINSs and CCOs straddles the lower end of the magnetar luminosities and is consistent with those of CCOs and XDINSs. Combining this multi-wavelength information may hint at the possibility that \xmmsrc~is a transitioning source in the process of evolving between the different sub-populations of neutron stars. This further highlights the need for regular monitoring of radio-quiet neutron stars to confirm this conjecture. The non-detection of any contemporaneous X-ray burst emission at 8-100 keV implies a radio-to-X-ray flux ratio $F_{\rm R}/F_{\rm X}>3\times10^{-8}$. The most luminous radio burst from SGR~1935+2154 and its contemporaneous X-ray burst resulted in a large $F_{\rm R}/F_{\rm X}\approx2\times10^{-5}$. If, conservatively, the radio counterpart emission energetics scale linearly with X-ray burst energy, the \textit{Fermi}-GBM non-detection cannot rule out a contemporaneous magnetar-like hard X-ray burst with $L_{\rm X}\sim10^{35-36}$~erg~s$^{-1}$ from \xmmsrc\ at the time of the radio burst. 

\subsection{The CCO radio pulsar link}
A second possibility is that \xmmsrc~ is a link between CCOs and radio pulsars. 
CCOs \citep[][]{Pavlov2004,deluca2017} are typically found in supernova remnants (SNRs) and are about a factor of 10 to 100 younger than the inferred ages for XDINS. Identifying an older SNR in the crowded region of the Carina nebula is difficult due to interactions of the strong winds of massive stars with their surroundings \citep{Townsley2011} and the extinction at optical and infra-red wavelengths. Our investigation of archival optical/IR data did not result in any obvious candidate for a SNR in the region. However, \cite{pires2015} reported on a runaway-star in the Carina nebula that may be associated with the birth of \xmmsrc~ $1.1-1.3 \times 10^4$\,yr ago. Such an age would be consistent with an evolved CCO.
The dipolar magnetic fields of CCOs are relatively weak, $B \sim 10^{10} - 10^{11}$\,G \citep{Gotthelf2007,Halpern2010} compared to XDINSs. 
A candidate X-ray absorption feature was also reported in the XMM-Newton spectrum of \xmmsrc~\citep{pires2012}. If that line is an electron cyclotron line, it implies a  magnetic field of 
$B_{\rm cyc} \sim 1.5 \times 10^{11}$\,G~\citep{pires2015}. A relatively weak magnetic field and the 18.6\,ms candidate spin period of \xmmsrc~are reminiscent of Calvera, a 59\,ms thermally X-ray emitting NS that was recently associated with a faint SNR \citep{arias2022, Rigoselli2024}. These two objects are currently the best candidates for the descendants of CCOs that otherwise have proven to be rather elusive \citep{Gotthelf2013,Luo2015}. The onset of radio emission is predicted for evolved CCOs~\citep{Luo2015,Ho2011}, representing the evolution from the CCO stage to the ``normal" radio pulsar stage. The observed radio burst of \xmmsrc\ may be the onset of the radio emission phase in \xmmsrc.

\subsection{Coherent radio emission from X-ray bright neutron stars}

\begin{figure}
    \centering
    \includegraphics[width=0.9\linewidth]{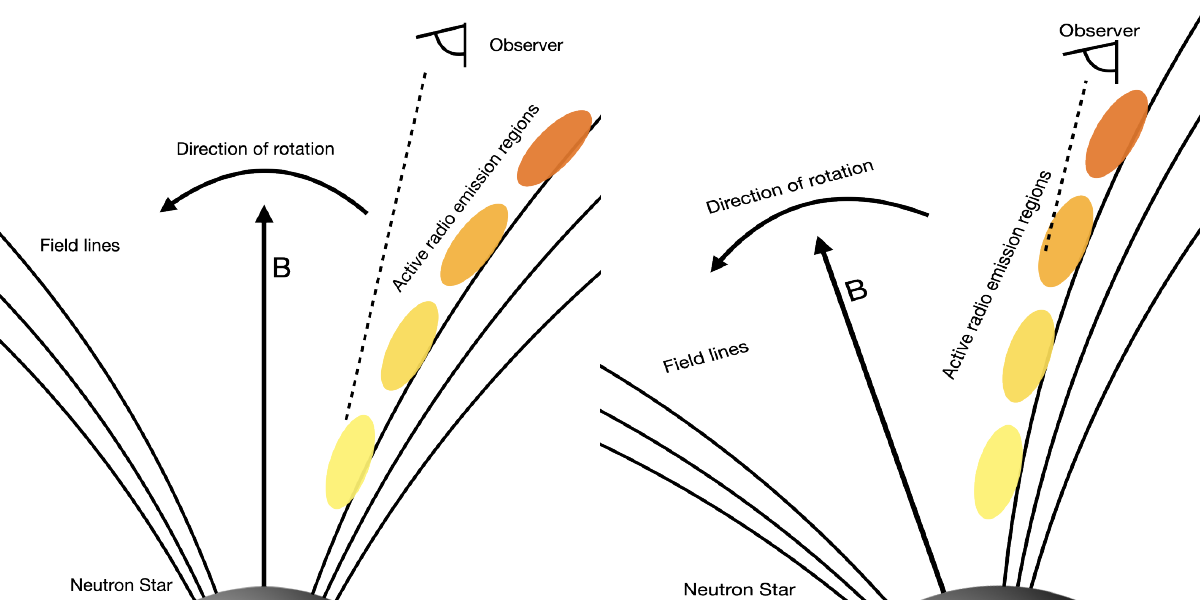}
    \caption{Schematic of a phenomenological model of the observed radio emission from \xmmsrc. Different active zones spawned at varied times and altitudes near a magnetic pole arrive at an observer from higher altitudes and more oblique angles at later times. This can explain the observed frequency evolution of emission and the flattening of the PPAs. The left panel denotes the emission seen at time $t_1$ while the right panel is the emission seen at time $t_2>t_1$ (observer frame).}
    \label{fig:toymodel}
\end{figure}

Until now, all XDINS and CCOs discovered in the Galaxy are radio quiet. The only connection comes from PSR~J0726-2612, a radio pulsar that also shares commonalities between its X-ray emission and the emission seen from XDINSs~\citep{rigoselli2019}. Historically, there have been many efforts to detect radio emission from these sources. The dearth of radio emission, whether transient or pulsed, led some to believe that the magnetic field in XDINSs is low enough to place them beyond the so-called death-line, the limit in spin-period and magnetic field beyond which radio emission is not possible under the current framework of rotationally powered emission theory. Based on observations of PSR~J0726$-$2612,~\cite{rigoselli2019} claimed that the lack of radio emission is a beaming effect rather than a genuine absence of emission from these sources. The radio burst reported in this work clearly shows that at least some of these X-ray isolated neutron stars have the ability to produce transient radio emission suggesting the possibility of a population of such bursts that are missed by radio telescopes due to their extreme episodic behaviour. The lack of any other radio burst before or after the event followed by a complete absence of any pulsed emission in follow-up radio observations or the archival MeerKAT observations (totalling 2.7 hours) suggests that the burst we discovered belongs to a category of transient radio emission resulting from transient episodes in the magnetosphere similar to the radio bursts seen in other magnetars. Assuming that the bursts are Poisson distributed in time, we compute the 99$\%$ confidence level upper-limit on the burst rate from the source of 2.4 bursts per hour above our sensitivity limit of 0.012~Jy. Furthermore, the shape of the burst is unique and does not share any similarities with typical radio pulses seen from other pulsars or RRATs. 
A simple phenomenological model can explain the shape and the observed dynamic spectrum, which shows shows multiple components that appear at steadily lower radio frequencies,  of the radio burst (Figure~\ref{fig:toymodel}). We represent each of these components to distinct emission regions possibly in motion along magnetic field lines near a pole. As the neutron star rotates across our line of sight, the observer intercepts emission from the different emission components decoupling at different heights along the magnetic field line, analogous to water blobs from a sprinkler. Different heights result in different components gradually arriving at lower radio frequency with time if we assume that the radio emission follows a pulsar-like radius-to-frequency mapping. The shape of the PPA can also be explained under this framework, as the impact parameter relative to the magnetic pole changes. The PPA changes rapidly for the leading component which could be explained by the lower altitude being closer to a pole compared to emission from oblique angles at higher altitudes leading to an apparent flattening of the PPA for trailing components. 
\section{Summary}
In summary, we report the serendipitous discovery a bright coherent radio transient from \xmmsrc, establishing its neutron star nature. The burst morphology is unique and deviates from what is observed from the radio pulse zoo of radio pulsars. The fast rise and an apparent exponential decay with emergence of several polarized components suggest a mechanism whereby a disturbance near the surface of a neutron star resulted in the creation of coherent emission observed near a magnetic pole. The lack of persistent pulsed radio emission before or after the event supports the transient nature of this event. This suggests that objects like \xmmsrc~are also capable of producing these sporadic radio bursts akin to the bursts seen from the magnetar SGR~J1935+2154~\citep{bochenek2020, SGR1935chime} and points to a common emission mechanism that operates over a range of magnetic fields and radio luminosities. The discovery of this burst also indicates that \xmmsrc\, may represent a unique evolutionary phase whereby the neutron stars are transitioning from being magnetically (magnetars) or thermally (CCOs/ XDINS) powered to rotationally powered neutron stars and may point to the onset of coherent radio emission commonly seen in pulsars.

\begin{acknowledgments}
This work was supported by a UKRI-STFC grant (SKA-NIPS, no. ST/Z510439/1). For the purpose of Open Access, the author has applied a CC BY public copyright licence to any Author Accepted Manuscript version arising from this submission. JT and BWS acknowledge funding from an STFC Consolidated grant. IPM acknowledges funding from an NWO Rubicon Fellowship, project number 019.221EN.019. M.C. acknowledges support of an Australian Research Council Discovery Early Career Research Award (project number DE220100819) funded by the Australian Government. The MeerTRAP collaboration acknowledges funding from the European Research Council under the European Union’s Horizon 2020 research and innovation programme (grant agreement No 694745). Z.W. acknowledges support by NASA under award numbers 80GSFC21M0002 and 80GSFC24M0006. The MeerTRAP collaboration would like to thank the MeerKAT Open time proposal teams for allowing MeerTRAP to observe commensally. We acknowledge the usage of TRAPUM infrastructure funded and installed by the Max-Planck-Institut f$\rm \ddot{u}$r Radioastronomie and the Max-Planck-Gesellschaft. We would like to thank SARAO science operations, CAM/CBF and operator teams for their time and effort invested in the observations. The MeerKAT telescope is operated by the South African Radio Astronomy Observatory, which is a facility of the National Research Foundation, an agency of the Department of Science and Innovation (DSI). 
This scientific work uses data obtained from telescopes within the Australia Telescope National Facility $\footnote{\url{https://ror.org/05qajvd42}}$ which is funded by the Australian Government for operation as a National Facility managed by CSIRO.  This work made use of data obtained with XMM-Newton, an ESA science mission with instruments and contributions directly funded by ESA Member States and NASA. This research has made use of data and software provided by the High Energy Astrophysics Science Archive Research Center (HEASARC), which is a service of the Astrophysics Science Division at NASA/GSFC and the High Energy Astrophysics Division of the Smithsonian Astrophysical Observatory. This work has made use of the NASA Astrophysics Data System.
\end{acknowledgments}

%

\vspace{5mm}
\facilities{XMM, NICER, Fermi, INTEGRAL, MeerKAT, Parkes}




\appendix

\section{X-ray observations}
\label{ref:xray}
We triggered prompt X-ray observations of \xmmsrc~with the Neutron star Interior Composition Explorer (NICER) on-board the International Space Station~\citep{gendreau2016}. The observations were taken on 2024-03-07 spanning a total of 260~seconds. The small integration times are attributed to larger Sun angle constraints on NICER to keep the background as low as possible. We also processed archival NICER data taken on the source from 2017 November 18 to 25 with observation IDs 1030090101 to 1030090108. We first generated level-2 data products using \textsc{NICERDAS} as part of the \textsc{HEASOFT} suite~\footnote{\url{https://heasarc.gsfc.nasa.gov/docs/software/heasoft/}}. The resulting lightcurves were checked for strong particle flaring at the harder X-ray energies. These flares were visually identified and removed to create clean Good Time Intervals (GTI) for the entire NICER dataset, resulting in a total integration time of 101.5~ks.
We also processed archival XMM-Newton observations taken in 2010 and 2012 with observation IDs 0650840101 
and 0691970101, and 91 and 88~ks exposures, respectively~\citep{pires2015}. The Observation Data Files (ODF) were 
processed using the \textsc{XMM-SAS} software suite~\footnote{\url{https://www.cosmos.esa.int/web/xmm-newton/sas}} 
version 21.0.0. We performed standard cleaning and filtering to both observations similarly (e.g., accepting only 
event patterns 0–12 and good X-ray events with FLAG=0). Furthermore, we excluded intervals of high background flaring 
activity as measured from source-free full field-of-view light curve in the energy range 10 to 15 keV binned at 
100~second resolution. Finally, we extracted source events from a circle centered at the best-fit PSF location as 
obtained with the SAS task \texttt{eregionanalyse}, having a radius of 30", which greatly increased the signal-to-noise 
compared to larger extraction regions. We then extracted all counts in the 0.5 to 2.0~keV range to look for any periodicity from the soft thermal X-ray emission from the source. We ran an H-test~\citep{buccheri1992} on all the timestamps in each of the 
GTIs to check for any period in the range of 0.01 to 100~seconds. We also ran a similar search in the 0.4 to 1.3~keV range similar to the search performed by~\citep{pires2012}. In both our searches, we did not find any peak above a 3$\sigma$ threshold for a 0.4--1.3~keV pulsed fraction limit of 14$\%$ after correcting for the number of trials. For higher periods, we detect much lower frequency power due to the observing cadence of \textit{NICER}, making the search for longer periods in these datasets challenging. We note that we did not search for any potential drift in pulse period from the presence of a period derivative given the short baseline of the observations.

Similarly, we ran a period search on the two \textit{XMM-Newton} observations separately. These data have already been thoroughly searched for periods in the range of 0.0114-100~seconds~\citep{pires2015}. We performed an H-test search for the same periods, yet we also included spins up to 5000 seconds. We did not detect any significant periodic signal in the energy range 0.2--5.0~keV and we derive a $3\sigma$ upper-limit of about 15\%, trial-corrected. Note the consistency in the upper-limits derived with XMM-Newton and NICER; while NICER provides a larger raw number of counts, the soft diffuse X-ray emission around the source substantially increases its background, which could more easily be eliminated with XMM-Newton imaging capabilities. We do not find the periodicity reported in~\cite{pires2012} in both the datasets. Given the low significance of the reported signal, our non-detection can be attributed to differing filtering criteria, source extraction region and/or energy filtering.

\subsection{Search for high-energy bursts}
\label{sec:fermigbm}
We searched for real-time hard X-ray triggers at the time of the radio burst detection from several high-energy, large field-of-view instruments, namely, \textit{Swift}-BAT, INTEGRAL-SPI, and \textit{Fermi}-GBM~\citep{winkler2003, meegan2009, krimm2003}. We do not find any plausible candidates, nor where there any reported through online notices (e.g., General Coordinates Network or Astronomers telegram). \xmmsrc\ was visible to GBM at the time of the radio burst (see Table~\ref{tab:mtp9g_prop}). Detectors n4 and n8 had favourable viewing angles ($<$60$^\circ$), hence, we used them to derive an upper-limit on a putative simultaneous hard X-ray emission. Under the assumption of a magnetar-like burst with 0.2~second duration and a thermal spectrum with $kT$=10~keV, we derive a $3\sigma$ upper-limit of $5.0\times10^{-8}$~erg~s$^{-1}$~cm$^{-2}$ in the 10-100 keV energy range, for a luminosity $L_{\rm X}\leq10^{37}$~erg~s$^{-1}$ at 1.5 kpc.
We also searched \textit{Fermi}-GBM for any weak, sub-threshold magnetar-like bursts from the direction of the source within a 2-month interval bracketing the radio burst, yet we do not find any. Moreover, \textit{Swift}-BAT, which is equipped with a coded mask instrument capable of arcminute-localization of the counterpart to any hard X-ray burst never detected any short, magnetar-like burst from the direction of \xmmsrc\, during its 20-year lifetime (or about 2 years of total exposure given the half-coded BAT field-of-view). Finally, we searched the targeted observations of the source, i.e., with NICER and XMM-Newton, for any short bursts. While these instruments cover a small time interval, they afford sensitivity two orders of magnitude deeper, ( i.e., $\sim5\times10^{-10}$~erg~s$^{-1}$~cm$^{-2}$) than large field of view instruments. The source did not exhibit any bursting events during these observations.

\section{Quasi-periodicity search}
\label{sec:qpsearch}
To determine the presence of a quasi-periodicity in the burst profile, we apply two separate techniques. First, we search for a periodicity in the ToAs of the burst subcomponents. Following \cite{ipm2023} and  \cite{chimefrb2022}, we fit the ToAs as a function of component number ($N$) to a linear function using a weighted least-squares minimization technique. By introducing gaps between the subcomponents, we find the subcomponent configuration that minimizes the reduced $\chi^2$ statistic, which resulted to be $N=(0,1,2,3,4,5,8,11)$. The slope of the linear function gives the candidate period $P_{\text{sc}}=4.15\pm0.08$~ms. We compute the significance of this candidate period by simulating random component spacings from a modified Poisson distribution with an exclusion parameter $\eta=0.2$ (minimal spacing between simulated components is $\eta P_{\text{sc}}$). We simulate $10^4$ distributions with 8 subcomponents each (7 component spacings), and by comparing their reduced $\chi^2$ statistic we find the significance of the candidate period $P_{\text{sc}}$ to be $2.1\sigma$. Hence, we conclude that there is no quasi-periodicity in the emission components.
Second, we search for a period in the power spectrum of the burst, which we generate with \textsc{Stingray} \citep{stingray}. The power spectrum can be well fitted to a power law with a white noise component. The lack of peaks in the power spectrum supports an absence of periodicity in the burst profile. 

\section{Localisation}
\label{sec:loc}

We imaged the full 300\,ms of voltage data to obtain a list of radio sources in the target field and matched them to the sources in the Rapid ASKAP Continuum Survey (RACS) catalogue \citep{McConnell20}. This resulted in 10 matches with the separations ranging from 1.5 to 4.7". We used these matched sources to solve for a transformation matrix to shift and rotate the MeerKAT sources to match the RACS source positions\footnote{The code for performing the astrometric correction can be found on GitHub: \url{https://github.com/AstroLaura/MeerKAT_Source_Matching}}, reducing the separations to 0.3--2.5\,arcsec with a median of 0.9\,arcsec. We then applied the transformation matrix to the image containing the burst and found the burst position to be RA = 10:46:08.6(1.8), Dec = $-$59:43:06.8(1.4). The uncertainty was determined by adding up three components in quadrature, a statistical component from source fitting in the image (1.2" in RA and 0.4" in Dec), the absolute astrometric uncertainty from the RACS positions (1" in both RA and Dec) and the median offset of the positions after the astrometric correction (0.9").

After accounting for astrometry, the position of the detected burst coincides with a known X-ray neutron star \xmmsrc~to within 1" suggesting an association with the compact object. To confirm the association, we computed the chance coincidence probability for the radio burst to be associated with \xmmsrc. Assuming a Poisson distribution of X-ray sources within the Carina complex, we estimate the number density of X-ray sources using the catalog presented in~\cite{broos2011}. Then, the probability of finding an X-ray source in the region within a radius $R$,
\begin{equation}
P({\rm R})_{\rm cc} = 1 - e^{-\rm \pi R^{2}~\rho},
\end{equation}
where $\rho$ is the X-ray source density. Using $\rm R\sim1$~arcsecond from the radio position uncertainty, we obtain $P({\rm R})_{\rm cc}\sim$~2.4$\times$10$^{-3}$. This rules out a spatial chance coincidence at a confidence level larger than 3$\sigma$. Another confirmation of the association comes from the expected distance of the radio burst. We use the publicly available electron density models of the Galaxy, namely \textsc{YMW16}~\citep{ymw16} and \textsc{NE2001}~\citep{ne2001} to estimate the distance to the source based on the measured DM. The distance estimate (1.5~kpc and 2.3~kpc respectively) is consistent (within model uncertainties) with the distance estimates published in the literature (1.5--2.5~kpc)~\citep{pires2012}.

\section{Radio Luminosity}
\label{sec:rl}
 To compute the burst luminosity, we first computed the flux density. Since the burst is not detected at the phase centre of the primary beam, we use the treatment presented in~\cite{jankowski2023} to correct for the various beam degradation effects. Hence, for a given pulse width $W$, the peak flux density of the burst,
\begin{equation}
S_{\rm peak}  = \beta~S/N~\eta~\frac{T_{\rm sys} + T_{\rm sky}}{G_{\rm ib}\sqrt{N_{p}~b_{\rm eff}~W}}~a_{\rm ib}^{-1},
\end{equation}
where $S/N$ is the signal-to-noise ratio, $T_{\rm sys}$ is the system temperature, $T_{\rm sky}$ is the sky temperature, $G_{\rm ib}$ is the telescope gain for the incoherent beam, $N_{p}$ is the number of polarizations to be summed, $\beta$ is the efficiency factor for digitisation, $b_{\rm eff}$ is the effective bandwidth, $\eta$ is the beamforming efficiency and $a_{\rm ib}$ is the beam degradation factor. Using typical values for MeerKAT from~\cite{jankowski2023} and $a_{\rm ib}$ = 0.73, we estimate $S_{\rm peak}$ of 2.66~Jy. Then the isotropic luminosity,
\begin{equation}
L_{\rm r} = ~9.52\times10^{13}~S_{\rm peak}\Delta \nu~4\pi D^{2}~\rm ergs~s^{-1},
\end{equation}
where $\Delta \nu$ is the bandwidth of the radio receiver in Hz, $D$ is the distance in pc and $S_{\rm peak}$ is the peak flux in Jy.




\bibliography{example}{}

\begin{thebibliography}{}
\expandafter\ifx\csname natexlab\endcsname\relax\def\natexlab#1{#1}\fi
\providecommand{\url}[1]{\href{#1}{#1}}
\providecommand{\dodoi}[1]{doi:~\href{http://doi.org/#1}{\nolinkurl{#1}}}
\providecommand{\doeprint}[1]{\href{http://ascl.net/#1}{\nolinkurl{http://ascl.net/#1}}}
\providecommand{\doarXiv}[1]{\href{https://arxiv.org/abs/#1}{\nolinkurl{https://arxiv.org/abs/#1}}}

\bibitem[{{Arias} {et~al.}(2022){Arias}, {Botteon}, {Bassa}, {van der Jagt}, {van Weeren}, {O'Sullivan}, {Bosschaart}, {Dullaart}, {Hardcastle}, {Hessels}, {Shimwell}, {Slob}, {Sturm}, {Tasse}, {Theijssen}, \& {Vink}}]{arias2022}
{Arias}, M., {Botteon}, A., {Bassa}, C.~G., {et~al.} 2022, \aap, 667, A71, \dodoi{10.1051/0004-6361/202244369}

\bibitem[{{Beniamini} {et~al.}(2019){Beniamini}, {Hotokezaka}, {van der Horst}, \& {Kouveliotou}}]{2019MNRAS.487.1426B}
{Beniamini}, P., {Hotokezaka}, K., {van der Horst}, A., \& {Kouveliotou}, C. 2019, \mnras, 487, 1426, \dodoi{10.1093/mnras/stz1391}

\bibitem[{{Beniamini} {et~al.}(2023){Beniamini}, {Wadiasingh}, {Hare}, {Rajwade}, {Younes}, \& {van der Horst}}]{beniamini2023}
{Beniamini}, P., {Wadiasingh}, Z., {Hare}, J., {et~al.} 2023, \mnras, 520, 1872, \dodoi{10.1093/mnras/stad208}

\bibitem[{{Bochenek} {et~al.}(2020){Bochenek}, {Ravi}, {Belov}, {Hallinan}, {Kocz}, {Kulkarni}, \& {McKenna}}]{bochenek2020}
{Bochenek}, C.~D., {Ravi}, V., {Belov}, K.~V., {et~al.} 2020, \nat, 587, 59, \dodoi{10.1038/s41586-020-2872-x}

\bibitem[{{Broos} {et~al.}(2011){Broos}, {Townsley}, {Feigelson}, {Getman}, {Garmire}, {Preibisch}, {Smith}, {Babler}, {Hodgkin}, {Indebetouw}, {Irwin}, {King}, {Lewis}, {Majewski}, {McCaughrean}, {Meade}, \& {Zinnecker}}]{broos2011}
{Broos}, P.~S., {Townsley}, L.~K., {Feigelson}, E.~D., {et~al.} 2011, \apjs, 194, 2, \dodoi{10.1088/0067-0049/194/1/2}

\bibitem[{{Buccheri}(1992)}]{buccheri1992}
{Buccheri}, R. 1992, in Statistical Challenges in Modern Astronomy, ed. E.~D. {Feigelson} \& G.~J. {Babu}, 381--409, \dodoi{10.1007/978-1-4613-9290-3_44}

\bibitem[{{Camilo} {et~al.}(2006){Camilo}, {Ransom}, {Halpern}, {Reynolds}, {Helfand}, {Zimmerman}, \& {Sarkissian}}]{camilo2006}
{Camilo}, F., {Ransom}, S.~M., {Halpern}, J.~P., {et~al.} 2006, \nat, 442, 892, \dodoi{10.1038/nature04986}

\bibitem[{{Camilo} {et~al.}(2018){Camilo}, {Scholz}, {Serylak}, {Buchner}, {Merryfield}, {Kaspi}, {Archibald}, {Bailes}, {Jameson}, {van Straten}, {Sarkissian}, {Reynolds}, {Johnston}, {Hobbs}, {Abbott}, {Adam}, {Adams}, {Alberts}, {Andreas}, {Asad}, {Baker}, {Baloyi}, {Bauermeister}, {Baxana}, {Bennett}, {Bernardi}, {Booisen}, {Booth}, {Botha}, {Boyana}, {Brederode}, {Burger}, {Cheetham}, {Conradie}, {Conradie}, {Davidson}, {De Bruin}, {de Swardt}, {de Villiers}, {de Villiers}, {de Villiers}, {de Villiers}, {De Waal}, {Dikgale}, {du Toit}, {du Toit}, {Esterhuyse}, {Fanaroff}, {Fataar}, {Foley}, {Foster}, {Fourie}, {Gamatham}, {Gatsi}, {Geschke}, {Goedhart}, {Grobler}, {Gumede}, {Hlakola}, {Hokwana}, {Hoorn}, {Horn}, {Horrell}, {Hugo}, {Isaacson}, {Jacobs}, {Jansen van Rensburg}, {Jonas}, {Jordaan}, {Joubert}, {Joubert}, {J{\'o}zsa}, {Julie}, {Julius}, {Kapp}, {Karastergiou}, {Karels}, {Kariseb}, {Karuppusamy}, {Kasper}, {Knox-Davies}, {Koch}, {Kotz{\'e}}, {Krebs}, {Kriek}, {Kriel}, {Kusel}, {Lamoor},
  {Lehmensiek}, {Liebenberg}, {Liebenberg}, {Lord}, {Lunsky}, {Mabombo}, {Macdonald}, {Macfarlane}, {Madisa}, {Mafhungo}, {Magnus}, {Magozore}, {Mahgoub}, {Main}, {Makhathini}, {Malan}, {Malgas}, {Manley}, {Manzini}, {Marais}, {Marais}, {Marais}, {Maree}, {Martens}, {Matshawule}, {Matthysen}, {Mauch}, {McNally}, {Merry}, {Millenaar}, {Mjikelo}, {Mkhabela}, {Mnyandu}, {Moeng}, {Mokone}, {Monama}, {Montshiwa}, {Moss}, {Mphego}, {New}, {Ngcebetsha}, {Ngoasheng}, {Niehaus}, {Ntuli}, {Nzama}, {Obies}, {Obrocka}, {Ockards}, {Olyn}, {Oozeer}, {Otto}, {Padayachee}, {Passmoor}, {Patel}, {Paula}, {Peens-Hough}, {Pholoholo}, {Prozesky}, {Rakoma}, {Ramaila}, {Rammala}, {Ramudzuli}, {Rasivhaga}, {Ratcliffe}, {Reader}, {Renil}, {Richter}, {Robyntjies}, {Rosekrans}, {Rust}, {Salie}, {Sambu}, {Schollar}, {Schwardt}, {Seranyane}, {Sethosa}, {Sharpe}, {Siebrits}, {Sirothia}, {Slabber}, {Smirnov}, {Smith}, {Sofeya}, {Songqumase}, {Spann}, {Stappers}, {Steyn}, {Steyn}, {Strong}, {Struthers}, {Stuart}, {Sunnylall}, {Swart},
  {Taljaard}, {Tasse}, {Taylor}, {Theron}, {Thondikulam}, {Thorat}, {Tiplady}, {Toruvanda}, {van Aardt}, {van Balla}, {van den Heever}, {van der Byl}, {van der Merwe}, {van der Merwe}, {van Niekerk}, {van Rooyen}, {van Staden}, {van Tonder}, \& {van Wyk}}]{camilo2018}
{Camilo}, F., {Scholz}, P., {Serylak}, M., {et~al.} 2018, \apj, 856, 180, \dodoi{10.3847/1538-4357/aab35a}

\bibitem[{{CHIME/FRB Collaboration} {et~al.}(2020){CHIME/FRB Collaboration}, {Andersen}, {Bandura}, {Bhardwaj}, {Bij}, {Boyce}, {Boyle}, {Brar}, {Cassanelli}, {Chawla}, {Chen}, {Cliche}, {Cook}, {Cubranic}, {Curtin}, {Denman}, {Dobbs}, {Dong}, {Fandino}, {Fonseca}, {Gaensler}, {Giri}, {Good}, {Halpern}, {Hill}, {Hinshaw}, {H{\"o}fer}, {Josephy}, {Kania}, {Kaspi}, {Landecker}, {Leung}, {Li}, {Lin}, {Masui}, {McKinven}, {Mena-Parra}, {Merryfield}, {Meyers}, {Michilli}, {Milutinovic}, {Mirhosseini}, {M{\"u}nchmeyer}, {Naidu}, {Newburgh}, {Ng}, {Patel}, {Pen}, {Pinsonneault-Marotte}, {Pleunis}, {Quine}, {Rafiei-Ravandi}, {Rahman}, {Ransom}, {Renard}, {Sanghavi}, {Scholz}, {Shaw}, {Shin}, {Siegel}, {Singh}, {Smegal}, {Smith}, {Stairs}, {Tan}, {Tendulkar}, {Tretyakov}, {Vanderlinde}, {Wang}, {Wulf}, \& {Zwaniga}}]{SGR1935chime}
{CHIME/FRB Collaboration}, {Andersen}, B.~C., {Bandura}, K.~M., {et~al.} 2020, \nat, 587, 54, \dodoi{10.1038/s41586-020-2863-y}

\bibitem[{{Chime/Frb Collaboration} {et~al.}(2022){Chime/Frb Collaboration}, {Bandura}, {Bhardwaj}, {Boyle}, {Brar}, {Breitman}, {Cassanelli}, {Chatterjee}, {Chawla}, {Cliche}, {Cubranic}, {Curtin}, {Deng}, {Dobbs}, {Dong}, {Fonseca}, {Gaensler}, {Giri}, {Good}, {Hill}, {Josephy}, {Kaczmarek}, {Kader}, {Kania}, {Kaspi}, {Leung}, {Li}, {Lin}, {Masui}, {McKinven}, {Mena-Parra}, {Merryfield}, {Meyers}, {Michilli}, {Naidu}, {Newburgh}, {Ng}, {Ordog}, {Patel}, {Pearlman}, {Pen}, {Petroff}, {Pleunis}, {Rafiei-Ravandi}, {Rahman}, {Ransom}, {Renard}, {Sanghavi}, {Scholz}, {Shaw}, {Shin}, {Siegel}, {Singh}, {Smith}, {Stairs}, {Tan}, {Tendulkar}, {Vanderlinde}, {Wiebe}, {Wulf}, \& {Zwaniga}}]{chimefrb2022}
{Chime/Frb Collaboration}, Bridget~C., A., {Bandura}, K., {Bhardwaj}, M., {et~al.} 2022, \nat, 607, 256, \dodoi{10.1038/s41586-022-04841-8}

\bibitem[{{Collazzi} {et~al.}(2015){Collazzi}, {Kouveliotou}, {van der Horst}, {Younes}, {Kaneko}, {G{\"o}{\u{g}}{\"u}{\c{s}}}, {Lin}, {Granot}, {Finger}, {Chaplin}, {Huppenkothen}, {Watts}, {von Kienlin}, {Baring}, {Gruber}, {Bhat}, {Gibby}, {Gehrels}, {McEnery}, {van der Klis}, \& {Wijers}}]{collazzi2011}
{Collazzi}, A.~C., {Kouveliotou}, C., {van der Horst}, A.~J., {et~al.} 2015, \apjs, 218, 11, \dodoi{10.1088/0067-0049/218/1/11}

\bibitem[{{Colpi} {et~al.}(2000){Colpi}, {Geppert}, \& {Page}}]{2000ApJ...529L..29C}
{Colpi}, M., {Geppert}, U., \& {Page}, D. 2000, \apjl, 529, L29, \dodoi{10.1086/312448}

\bibitem[{{Cooper} \& {Wadiasingh}(2024)}]{cooper2024}
{Cooper}, A.~J., \& {Wadiasingh}, Z. 2024, \mnras, 533, 2133, \dodoi{10.1093/mnras/stae1813}

\bibitem[{{Cordes} \& {Lazio}(2002)}]{ne2001}
{Cordes}, J.~M., \& {Lazio}, T.~J.~W. 2002, arXiv e-prints, astro.
\newblock \doarXiv{astro-ph/0207156}

\bibitem[{{Dall'Osso} {et~al.}(2012){Dall'Osso}, {Granot}, \& {Piran}}]{2012MNRAS.422.2878D}
{Dall'Osso}, S., {Granot}, J., \& {Piran}, T. 2012, \mnras, 422, 2878, \dodoi{10.1111/j.1365-2966.2012.20612.x}

\bibitem[{{De Luca}(2017)}]{deluca2017}
{De Luca}, A. 2017, in Journal of Physics Conference Series, Vol. 932, Journal of Physics Conference Series (IOP), 012006, \dodoi{10.1088/1742-6596/932/1/012006}

\bibitem[{{Esposito} {et~al.}(2020){Esposito}, {Rea}, {Borghese}, {Coti Zelati}, {Vigan{\`o}}, {Israel}, {Tiengo}, {Ridolfi}, {Possenti}, {Burgay}, {G{\"o}tz}, {Pintore}, {Stella}, {Dehman}, {Ronchi}, {Campana}, {Garcia-Garcia}, {Graber}, {Mereghetti}, {Perna}, {Rodr{\'\i}guez Castillo}, {Turolla}, \& {Zane}}]{esposito2020}
{Esposito}, P., {Rea}, N., {Borghese}, A., {et~al.} 2020, \apjl, 896, L30, \dodoi{10.3847/2041-8213/ab9742}

\bibitem[{{Gendreau} {et~al.}(2016){Gendreau}, {Arzoumanian}, {Adkins}, {Albert}, {Anders}, {Aylward}, {Baker}, {Balsamo}, {Bamford}, {Benegalrao}, {Berry}, {Bhalwani}, {Black}, {Blaurock}, {Bronke}, {Brown}, {Budinoff}, {Cantwell}, {Cazeau}, {Chen}, {Clement}, {Colangelo}, {Coleman}, {Coopersmith}, {Dehaven}, {Doty}, {Egan}, {Enoto}, {Fan}, {Ferro}, {Foster}, {Galassi}, {Gallo}, {Green}, {Grosh}, {Ha}, {Hasouneh}, {Heefner}, {Hestnes}, {Hoge}, {Jacobs}, {J{\o}rgensen}, {Kaiser}, {Kellogg}, {Kenyon}, {Koenecke}, {Kozon}, {LaMarr}, {Lambertson}, {Larson}, {Lentine}, {Lewis}, {Lilly}, {Liu}, {Malonis}, {Manthripragada}, {Markwardt}, {Matonak}, {Mcginnis}, {Miller}, {Mitchell}, {Mitchell}, {Mohammed}, {Monroe}, {Montt de Garcia}, {Mul{\'e}}, {Nagao}, {Ngo}, {Norris}, {Norwood}, {Novotka}, {Okajima}, {Olsen}, {Onyeachu}, {Orosco}, {Peterson}, {Pevear}, {Pham}, {Pollard}, {Pope}, {Powers}, {Powers}, {Price}, {Prigozhin}, {Ramirez}, {Reid}, {Remillard}, {Rogstad}, {Rosecrans}, {Rowe}, {Sager}, {Sanders},
  {Savadkin}, {Saylor}, {Schaeffer}, {Schweiss}, {Semper}, {Serlemitsos}, {Shackelford}, {Soong}, {Struebel}, {Vezie}, {Villasenor}, {Winternitz}, {Wofford}, {Wright}, {Yang}, \& {Yu}}]{gendreau2016}
{Gendreau}, K.~C., {Arzoumanian}, Z., {Adkins}, P.~W., {et~al.} 2016, in Society of Photo-Optical Instrumentation Engineers (SPIE) Conference Series, Vol. 9905, Space Telescopes and Instrumentation 2016: Ultraviolet to Gamma Ray, ed. J.-W.~A. {den Herder}, T.~{Takahashi}, \& M.~{Bautz}, 99051H, \dodoi{10.1117/12.2231304}

\bibitem[{{Gotthelf} \& {Halpern}(2007)}]{Gotthelf2007}
{Gotthelf}, E.~V., \& {Halpern}, J.~P. 2007, \apjl, 664, L35, \dodoi{10.1086/520637}

\bibitem[{{Gotthelf} {et~al.}(2013){Gotthelf}, {Halpern}, {Allen}, \& {Knispel}}]{Gotthelf2013}
{Gotthelf}, E.~V., {Halpern}, J.~P., {Allen}, B., \& {Knispel}, B. 2013, \apj, 773, 141, \dodoi{10.1088/0004-637X/773/2/141}

\bibitem[{{Haberl}(2007)}]{haberl2007}
{Haberl}, F. 2007, \apss, 308, 181, \dodoi{10.1007/s10509-007-9342-x}

\bibitem[{{Halpern} \& {Gotthelf}(2010)}]{Halpern2010}
{Halpern}, J.~P., \& {Gotthelf}, E.~V. 2010, \apj, 709, 436, \dodoi{10.1088/0004-637X/709/1/436}

\bibitem[{{Hamaguchi} {et~al.}(2009){Hamaguchi}, {Corcoran}, {Ezoe}, {Townsley}, {Broos}, {Gruendl}, {Vaidya}, {White}, {Strohmayer}, {Petre}, \& {Chu}}]{Hamaguchi2009}
{Hamaguchi}, K., {Corcoran}, M.~F., {Ezoe}, Y., {et~al.} 2009, \apjl, 695, L4, \dodoi{10.1088/0004-637X/695/1/L4}

\bibitem[{{Ho}(2011)}]{Ho2011}
{Ho}, W. C.~G. 2011, \mnras, 414, 2567, \dodoi{10.1111/j.1365-2966.2011.18576.x}

\bibitem[{{Hobbs} {et~al.}(2020){Hobbs}, {Manchester}, {Dunning}, {Jameson}, {Roberts}, {George}, {Green}, {Tuthill}, {Toomey}, {Kaczmarek}, {Mader}, {Marquarding}, {Ahmed}, {Amy}, {Bailes}, {Beresford}, {Bhat}, {Bock}, {Bourne}, {Bowen}, {Brothers}, {Cameron}, {Carretti}, {Carter}, {Castillo}, {Chekkala}, {Cheng}, {Chung}, {Craig}, {Dai}, {Dawson}, {Dempsey}, {Doherty}, {Dong}, {Edwards}, {Ergesh}, {Gao}, {Han}, {Hayman}, {Indermuehle}, {Jeganathan}, {Johnston}, {Kanoniuk}, {Kesteven}, {Kramer}, {Leach}, {Mcintyre}, {Moss}, {Os{\l}owski}, {Phillips}, {Pope}, {Preisig}, {Price}, {Reeves}, {Reilly}, {Reynolds}, {Robishaw}, {Roush}, {Ruckley}, {Sadler}, {Sarkissian}, {Severs}, {Shannon}, {Smart}, {Smith}, {Smith}, {Sobey}, {Staveley-Smith}, {Tzioumis}, {van Straten}, {Wang}, {Wen}, \& {Whiting}}]{Hobbs2020}
{Hobbs}, G., {Manchester}, R.~N., {Dunning}, A., {et~al.} 2020, \pasa, 37, e012, \dodoi{10.1017/pasa.2020.2}

\bibitem[{{Huppenkothen} {et~al.}(2019){Huppenkothen}, {Bachetti}, {Stevens}, {Migliari}, {Balm}, {Hammad}, {Khan}, {Mishra}, {Rashid}, {Sharma}, {Martinez Ribeiro}, \& {Valles Blanco}}]{stingray}
{Huppenkothen}, D., {Bachetti}, M., {Stevens}, A.~L., {et~al.} 2019, \apj, 881, 39, \dodoi{10.3847/1538-4357/ab258d}

\bibitem[{Igoshev {et~al.}(2025)Igoshev, Barrère, Raynaud, Guilet, Wood, \& Hollerbach}]{igoshev2025protoneutron}
Igoshev, A., Barrère, P., Raynaud, R., {et~al.} 2025, From proto-neutron star dynamo to low-field magnetars.
\newblock \doarXiv{2501.04768}

\bibitem[{{Israel} {et~al.}(2021){Israel}, {Burgay}, {Rea}, {Esposito}, {Possenti}, {Dall'Osso}, {Stella}, {Pilia}, {Tiengo}, {Ridnaia}, {Lien}, {Frederiks}, \& {Bernardini}}]{israel2021}
{Israel}, G.~L., {Burgay}, M., {Rea}, N., {et~al.} 2021, \apj, 907, 7, \dodoi{10.3847/1538-4357/abca95}

\bibitem[{{Jankowski} {et~al.}(2023){Jankowski}, {Bezuidenhout}, {Caleb}, {Driessen}, {Malenta}, {Morello}, {Rajwade}, {Sanidas}, {Stappers}, {Surnis}, {Barr}, {Chen}, {Kramer}, {Wu}, {Buchner}, {Serylak}, \& {Prochaska}}]{jankowski2023}
{Jankowski}, F., {Bezuidenhout}, M.~C., {Caleb}, M., {et~al.} 2023, \mnras, 524, 4275, \dodoi{10.1093/mnras/stad2041}

\bibitem[{{Jensen} {et~al.}(2003){Jensen}, {Clausen}, {Cassi}, {Ravera}, {Janin}, {Winkler}, \& {Much}}]{jensen2003}
{Jensen}, P.~L., {Clausen}, K., {Cassi}, C., {et~al.} 2003, \aap, 411, L7, \dodoi{10.1051/0004-6361:20031173}

\bibitem[{{Kaplan} \& {van Kerkwijk}(2009)}]{Kaplan2009}
{Kaplan}, D.~L., \& {van Kerkwijk}, M.~H. 2009, \apj, 705, 798, \dodoi{10.1088/0004-637X/705/1/798}

\bibitem[{{Kaspi}(2010)}]{Kaspi2010}
{Kaspi}, V.~M. 2010, Proceedings of the National Academy of Science, 107, 7147, \dodoi{10.1073/pnas.1000812107}

\bibitem[{{Kaspi} \& {Beloborodov}(2017)}]{kaspi2017}
{Kaspi}, V.~M., \& {Beloborodov}, A.~M. 2017, \araa, 55, 261, \dodoi{10.1146/annurev-astro-081915-023329}

\bibitem[{{Kirsten} {et~al.}(2021){Kirsten}, {Snelders}, {Jenkins}, {Nimmo}, {van den Eijnden}, {Hessels}, {Gawro{\'n}ski}, \& {Yang}}]{kirsten2021}
{Kirsten}, F., {Snelders}, M.~P., {Jenkins}, M., {et~al.} 2021, Nature Astronomy, 5, 414, \dodoi{10.1038/s41550-020-01246-3}

\bibitem[{{Kondratiev} {et~al.}(2009){Kondratiev}, {McLaughlin}, {Lorimer}, {Burgay}, {Possenti}, {Turolla}, {Popov}, \& {Zane}}]{kondratiev2009}
{Kondratiev}, V.~I., {McLaughlin}, M.~A., {Lorimer}, D.~R., {et~al.} 2009, \apj, 702, 692, \dodoi{10.1088/0004-637X/702/1/692}

\bibitem[{{Kramer} {et~al.}(2024){Kramer}, {Liu}, {Desvignes}, {Karuppusamy}, \& {Stappers}}]{kramer2024}
{Kramer}, M., {Liu}, K., {Desvignes}, G., {Karuppusamy}, R., \& {Stappers}, B.~W. 2024, Nature Astronomy, 8, 230, \dodoi{10.1038/s41550-023-02125-3}

\bibitem[{{Kramer} {et~al.}(2006){Kramer}, {Lyne}, {O'Brien}, {Jordan}, \& {Lorimer}}]{kramer2006}
{Kramer}, M., {Lyne}, A.~G., {O'Brien}, J.~T., {Jordan}, C.~A., \& {Lorimer}, D.~R. 2006, Science, 312, 549, \dodoi{10.1126/science.1124060}

\bibitem[{{Krimm} {et~al.}(2003){Krimm}, {Barthelmy}, {Gehrels}, {Parsons}, {Tueller}, {Fenimore}, {Palmer}, {Markwardt}, {Belloni}, {Banat}, {Dean}, \& {Willis}}]{krimm2003}
{Krimm}, H.~A., {Barthelmy}, S.~D., {Gehrels}, N., {et~al.} 2003, in AAS/High Energy Astrophysics Division, Vol.~7, AAS/High Energy Astrophysics Division \#7, 16.14

\bibitem[{{Krimm} {et~al.}(2013){Krimm}, {Holland}, {Corbet}, {Pearlman}, {Romano}, {Kennea}, {Bloom}, {Barthelmy}, {Baumgartner}, {Cummings}, {Gehrels}, {Lien}, {Markwardt}, {Palmer}, {Sakamoto}, {Stamatikos}, \& {Ukwatta}}]{krimm2013}
{Krimm}, H.~A., {Holland}, S.~T., {Corbet}, R.~H.~D., {et~al.} 2013, \apjs, 209, 14, \dodoi{10.1088/0067-0049/209/1/14}

\bibitem[{{Lu} {et~al.}(2024){Lu}, {Zhou}, {Wang}, {Shao}, {Li}, {Vink}, {Li}, \& {Chen}}]{lu2024}
{Lu}, W.-J., {Zhou}, P., {Wang}, P., {et~al.} 2024, \apj, 963, 151, \dodoi{10.3847/1538-4357/ad27cf}

\bibitem[{{Luo} {et~al.}(2015){Luo}, {Ng}, {Ho}, {Bogdanov}, {Kaspi}, \& {He}}]{Luo2015}
{Luo}, J., {Ng}, C.~Y., {Ho}, W.~C.~G., {et~al.} 2015, \apj, 808, 130, \dodoi{10.1088/0004-637X/808/2/130}

\bibitem[{{Malofeev} {et~al.}(2006){Malofeev}, {Malov}, {Teplykh}, {Logvinenko}, {Litvinov}, \& {Popov}}]{malofeev2006}
{Malofeev}, V.~M., {Malov}, O.~I., {Teplykh}, D.~A., {et~al.} 2006, The Astronomer's Telegram, 798, 1

\bibitem[{{Malov} \& {Timirkeeva}(2018)}]{malov2018}
{Malov}, I.~F., \& {Timirkeeva}, M.~A. 2018, in Modern Star Astronomy, Vol.~1, 220--223, \dodoi{10.31361/eaas.2018-1.048}

\bibitem[{{Manchester} {et~al.}(2005){Manchester}, {Hobbs}, {Teoh}, \& {Hobbs}}]{psrcat}
{Manchester}, R.~N., {Hobbs}, G.~B., {Teoh}, A., \& {Hobbs}, M. 2005, \aj, 129, 1993, \dodoi{10.1086/428488}

\bibitem[{{McConnell} {et~al.}(2020){McConnell}, {Hale}, {Lenc}, {Banfield}, {Heald}, {Hotan}, {Leung}, {Moss}, {Murphy}, {O'Brien}, {Pritchard}, {Raja}, {Sadler}, {Stewart}, {Thomson}, {Whiting}, {Allison}, {Amy}, {Anderson}, {Ball}, {Bannister}, {Bell}, {Bock}, {Bolton}, {Bunton}, {Chippendale}, {Collier}, {Cooray}, {Cornwell}, {Diamond}, {Edwards}, {Gupta}, {Hayman}, {Heywood}, {Jackson}, {Koribalski}, {Lee-Waddell}, {McClure-Griffiths}, {Ng}, {Norris}, {Phillips}, {Reynolds}, {Roxby}, {Schinckel}, {Shields}, {Tremblay}, {Tzioumis}, {Voronkov}, \& {Westmeier}}]{McConnell20}
{McConnell}, D., {Hale}, C.~L., {Lenc}, E., {et~al.} 2020, \pasa, 37, e048, \dodoi{10.1017/pasa.2020.41}

\bibitem[{{McLaughlin} {et~al.}(2006){McLaughlin}, {Lyne}, {Lorimer}, {Kramer}, {Faulkner}, {Manchester}, {Cordes}, {Camilo}, {Possenti}, {Stairs}, {Hobbs}, {D'Amico}, {Burgay}, \& {O'Brien}}]{mclaughlin2006}
{McLaughlin}, M.~A., {Lyne}, A.~G., {Lorimer}, D.~R., {et~al.} 2006, \nat, 439, 817, \dodoi{10.1038/nature04440}

\bibitem[{{Meegan} {et~al.}(2009){Meegan}, {Lichti}, {Bhat}, {Bissaldi}, {Briggs}, {Connaughton}, {Diehl}, {Fishman}, {Greiner}, {Hoover}, {van der Horst}, {von Kienlin}, {Kippen}, {Kouveliotou}, {McBreen}, {Paciesas}, {Preece}, {Steinle}, {Wallace}, {Wilson}, \& {Wilson-Hodge}}]{meegan2009}
{Meegan}, C., {Lichti}, G., {Bhat}, P.~N., {et~al.} 2009, \apj, 702, 791, \dodoi{10.1088/0004-637X/702/1/791}

\bibitem[{{Men} \& {Barr}(2024)}]{Men24}
{Men}, Y., \& {Barr}, E. 2024, \aap, 683, A183, \dodoi{10.1051/0004-6361/202348247}

\bibitem[{{Morello} {et~al.}(2020){Morello}, {Barr}, {Stappers}, {Keane}, \& {Lyne}}]{Morello20}
{Morello}, V., {Barr}, E.~D., {Stappers}, B.~W., {Keane}, E.~F., \& {Lyne}, A.~G. 2020, \mnras, 497, 4654, \dodoi{10.1093/mnras/staa2291}

\bibitem[{{Oswald} {et~al.}(2021){Oswald}, {Karastergiou}, {Posselt}, {Johnston}, {Bailes}, {Buchner}, {Geyer}, {Keith}, {Kramer}, {Parthasarathy}, {Reardon}, {Serylak}, {Shannon}, {Spiewak}, {van Straten}, \& {Venkatraman Krishnan}}]{Oswald2021}
{Oswald}, L.~S., {Karastergiou}, A., {Posselt}, B., {et~al.} 2021, \mnras, 504, 1115, \dodoi{10.1093/mnras/stab980}

\bibitem[{{Pastor-Marazuela} {et~al.}(2023{\natexlab{a}}){Pastor-Marazuela}, {Straal}, {van Leeuwen}, \& {Kondratiev}}]{ipm2023b}
{Pastor-Marazuela}, I., {Straal}, S.~M., {van Leeuwen}, J., \& {Kondratiev}, V.~I. 2023{\natexlab{a}}, \aap, 672, A151, \dodoi{10.1051/0004-6361/202245214}

\bibitem[{{Pastor-Marazuela} {et~al.}(2023{\natexlab{b}}){Pastor-Marazuela}, {van Leeuwen}, {Bilous}, {Connor}, {Maan}, {Oostrum}, {Petroff}, {Straal}, {Vohl}, {Adams}, {Adebahr}, {Attema}, {Boersma}, {van den Brink}, {van Cappellen}, {Coolen}, {Damstra}, {D{\'e}nes}, {Hess}, {van der Hulst}, {Hut}, {Kutkin}, {Marcel Loose}, {Lucero}, {Mika}, {Moss}, {Mulder}, {Norden}, {Oosterloo}, {Rajwade}, {van der Schuur}, {Sclocco}, {Smits}, \& {Ziemke}}]{ipm2023}
{Pastor-Marazuela}, I., {van Leeuwen}, J., {Bilous}, A., {et~al.} 2023{\natexlab{b}}, \aap, 678, A149, \dodoi{10.1051/0004-6361/202243339}

\bibitem[{{Pavlov} {et~al.}(2004){Pavlov}, {Sanwal}, \& {Teter}}]{Pavlov2004}
{Pavlov}, G.~G., {Sanwal}, D., \& {Teter}, M.~A. 2004, in IAU Symposium, Vol. 218, Young Neutron Stars and Their Environments, ed. F.~{Camilo} \& B.~M. {Gaensler}, 239, \dodoi{10.48550/arXiv.astro-ph/0311526}

\bibitem[{{Pires} {et~al.}(2015){Pires}, {Motch}, {Turolla}, {Popov}, {Schwope}, \& {Treves}}]{pires2015}
{Pires}, A.~M., {Motch}, C., {Turolla}, R., {et~al.} 2015, \aap, 583, A117, \dodoi{10.1051/0004-6361/201526436}

\bibitem[{{Pires} {et~al.}(2012){Pires}, {Motch}, {Turolla}, {Schwope}, {Pilia}, {Treves}, {Popov}, \& {Janot-Pacheco}}]{pires2012}
---. 2012, \aap, 544, A17, \dodoi{10.1051/0004-6361/201219161}

\bibitem[{{Prinz} \& {Becker}(2015)}]{prinz2015}
{Prinz}, T., \& {Becker}, W. 2015, arXiv e-prints, arXiv:1511.07713, \dodoi{10.48550/arXiv.1511.07713}

\bibitem[{{Rajwade} {et~al.}(2022){Rajwade}, {Bezuidenhout}, {Caleb}, {Driessen}, {Jankowski}, {Malenta}, {Morello}, {Sanidas}, {Stappers}, {Surnis}, {Barr}, {Chen}, {Kramer}, {Wu}, {Buchner}, {Serylak}, {Combes}, {Fong}, {Gupta}, {Jagannathan}, {Kilpatrick}, {Krogager}, {Noterdaeme}, {N{\'u}nẽz}, {Prochaska}, {Srianand}, \& {Tejos}}]{rajwade2022}
{Rajwade}, K.~M., {Bezuidenhout}, M.~C., {Caleb}, M., {et~al.} 2022, \mnras, 514, 1961, \dodoi{10.1093/mnras/stac1450}

\bibitem[{{Rajwade} {et~al.}(2024){Rajwade}, {Driessen}, {Barr}, {Pastor-Marazuela}, {Berezina}, {Jankowski}, {Muller}, {Kahinga}, {Stappers}, {Bezuidenhout}, {Caleb}, {Deller}, {Fong}, {Gordon}, {Kramer}, {Malenta}, {Morello}, {Prochaska}, {Sanidas}, {Surnis}, {Tejos}, \& {Wagner}}]{rajwade2024}
{Rajwade}, K.~M., {Driessen}, L.~N., {Barr}, E.~D., {et~al.} 2024, \mnras, \dodoi{10.1093/mnras/stae1652}

\bibitem[{{Rea} {et~al.}(2016){Rea}, {Borghese}, {Esposito}, {Coti Zelati}, {Bachetti}, {Israel}, \& {De Luca}}]{rea2016}
{Rea}, N., {Borghese}, A., {Esposito}, P., {et~al.} 2016, \apjl, 828, L13, \dodoi{10.3847/2041-8205/828/1/L13}

\bibitem[{{Rigoselli} {et~al.}(2024){Rigoselli}, {Mereghetti}, {Halpern}, {Gotthelf}, \& {Bassa}}]{Rigoselli2024}
{Rigoselli}, M., {Mereghetti}, S., {Halpern}, J.~P., {Gotthelf}, E.~V., \& {Bassa}, C.~G. 2024, \apj, 976, 228, \dodoi{10.3847/1538-4357/ad8cd6}

\bibitem[{{Rigoselli} {et~al.}(2019){Rigoselli}, {Mereghetti}, {Suleimanov}, {Potekhin}, {Turolla}, {Taverna}, \& {Pintore}}]{rigoselli2019}
{Rigoselli}, M., {Mereghetti}, S., {Suleimanov}, V., {et~al.} 2019, \aap, 627, A69, \dodoi{10.1051/0004-6361/201935485}

\bibitem[{{Russell} {et~al.}(2024){Russell}, {Degenaar}, {van den Eijnden}, {Maccarone}, {Tetarenko}, {S{\'a}nchez-Fern{\'a}ndez}, {Miller-Jones}, {Kuulkers}, \& {Del Santo}}]{russell2024}
{Russell}, T.~D., {Degenaar}, N., {van den Eijnden}, J., {et~al.} 2024, \nat, 627, 763, \dodoi{10.1038/s41586-024-07133-5}

\bibitem[{{Townsley} {et~al.}(2011){Townsley}, {Broos}, {Chu}, {Gruendl}, {Oey}, \& {Pittard}}]{Townsley2011}
{Townsley}, L.~K., {Broos}, P.~S., {Chu}, Y.-H., {et~al.} 2011, \apjs, 194, 16, \dodoi{10.1088/0067-0049/194/1/16}

\bibitem[{{Vigan{\`o}} {et~al.}(2013){Vigan{\`o}}, {Rea}, {Pons}, {Perna}, {Aguilera}, \& {Miralles}}]{vigano2013}
{Vigan{\`o}}, D., {Rea}, N., {Pons}, J.~A., {et~al.} 2013, \mnras, 434, 123, \dodoi{10.1093/mnras/stt1008}

\bibitem[{{Wadiasingh} {et~al.}(2020){Wadiasingh}, {Beniamini}, {Timokhin}, {Baring}, {van der Horst}, {Harding}, \& {Kazanas}}]{wadiasingh2020}
{Wadiasingh}, Z., {Beniamini}, P., {Timokhin}, A., {et~al.} 2020, \apj, 891, 82, \dodoi{10.3847/1538-4357/ab6d69}

\bibitem[{{Wadiasingh} \& {Chirenti}(2020)}]{wadiasingh2020b}
{Wadiasingh}, Z., \& {Chirenti}, C. 2020, \apjl, 903, L38, \dodoi{10.3847/2041-8213/abc562}

\bibitem[{{Wadiasingh} \& {Timokhin}(2019)}]{2019ApJ...879....4W}
{Wadiasingh}, Z., \& {Timokhin}, A. 2019, \apj, 879, 4, \dodoi{10.3847/1538-4357/ab2240}

\bibitem[{{Winkler} {et~al.}(2003){Winkler}, {Courvoisier}, {Di Cocco}, {Gehrels}, {Gim{\'e}nez}, {Grebenev}, {Hermsen}, {Mas-Hesse}, {Lebrun}, {Lund}, {Palumbo}, {Paul}, {Roques}, {Schnopper}, {Sch{\"o}nfelder}, {Sunyaev}, {Teegarden}, {Ubertini}, {Vedrenne}, \& {Dean}}]{winkler2003}
{Winkler}, C., {Courvoisier}, T.~J.~L., {Di Cocco}, G., {et~al.} 2003, \aap, 411, L1, \dodoi{10.1051/0004-6361:20031288}

\bibitem[{{Yao} {et~al.}(2017){Yao}, {Manchester}, \& {Wang}}]{ymw16}
{Yao}, J.~M., {Manchester}, R.~N., \& {Wang}, N. 2017, \apj, 835, 29, \dodoi{10.3847/1538-4357/835/1/29}

\bibitem[{{Yoneyama} {et~al.}(2019){Yoneyama}, {Hayashida}, {Nakajima}, \& {Matsumoto}}]{yoneyama2019}
{Yoneyama}, T., {Hayashida}, K., {Nakajima}, H., \& {Matsumoto}, H. 2019, \pasj, 71, 17, \dodoi{10.1093/pasj/psy135}

\bibitem[{{Zhu} {et~al.}(2023){Zhu}, {Xu}, {Zhou}, {Lin}, {Wang}, {Wang}, {Zhang}, {Niu}, {Chen}, {Li}, {Meng}, {Lee}, {Zhang}, {Feng}, {Ge}, {G{\"o}{\u{g}}{\"u}{\c{s}}}, {Guan}, {Han}, {Jiang}, {Jiang}, {Kouveliotou}, {Li}, {Miao}, {Miao}, {Men}, {Niu}, {Wang}, {Wang}, {Xu}, {Xu}, {Xue}, {Yang}, {Yu}, {Yuan}, {Yue}, {Zhang}, \& {Zhang}}]{2023SciA....9F6198Z}
{Zhu}, W., {Xu}, H., {Zhou}, D., {et~al.} 2023, Science Advances, 9, eadf6198, \dodoi{10.1126/sciadv.adf6198}

\end{thebibliography}
\bibliographystyle{aasjournal}



\end{document}